\documentstyle[12pt]{article}
\begin{document}

\baselineskip 24pt

\begin{center}

{\Large\bf
Interface roughening with a time-varying external driving force
}

\bigskip

E. Hern\'andez-Garc\'\i a,$^1$ T. Ala-Nissila,$^2$ and Martin Grant$^3$

\bigskip

{\it
$^1$Departament de F\'\i sica, Universitat de les Illes Balears,\\
E--07071, Palma de Mallorca, Spain
}

\bigskip

{\it
$^2$University of Helsinki,
Research Institute for Theoretical Physics, \\Siltavuorenpenger 20 C,
SF--00170 Helsinki, Finland
}

\bigskip

{\it
$^3$Centre for the Physics of Materials, Physics Department, \\
Rutherford Building, McGill University, 3600 rue University, \\
Montr\'eal, Qu\'ebec, Canada H3A 2T8
}

\end{center}

\vfill

We present a theoretical and numerical investigation of the effect of a
time-varying external driving force on interface growth.  First, we
derive
a relation between the roughening exponents which comes from a
generalized Galilean invariance, showing how the critical dimension of
the model is tunable with the external field.  We further conjecture
results for the exponents in two dimensions, and find consistency with data
obtained through simulations of two models we expect
to be in the same universality class. Finally, we discuss how our results
can be investigated experimentally.

\vfill

{\noindent \footnotesize
Submitted to Europhysics Letters.
}

\newpage

Open systems pose among the most challenging problems in
condensed-matter physics.  The transient dynamics and rich steady-state
behavior of systems ranging from dendrites to convective fluid cells
provide a unique testing ground for the ideas and methods of
nonequilibrium physics.  Many such problems involve pattern formation
where interfaces are present.  A particularly well-defined problem of
this kind is the roughening  of a growing interface \cite{Krug}.
A great deal of
theoretical progress has been made on this problem by analytic and
numerical methods, following the now-classic work of Kardar, Parisi,
and Zhang (KPZ) \cite{Kardar}.  They proposed a nonlinear differential
equation to model the height $h$ of a growing interface driven by an
external flux of particles:
\begin{equation}
\label{kpz}
 \frac{\partial h}{\partial t} =
  \nu \frac{\partial^2 h}{\partial {\vec x}^2}  +
  \frac{\lambda}{2}\left(\frac{\partial h}{\partial {\vec x}} \right)^2
+ \eta,
\end{equation}
where $\nu$ and $\lambda$ are constants, and $\eta$ is a random noise
which is assumed to satisfy Gaussian statistics with $ \langle \eta
(\vec{x},t)\eta (\vec{x}',t')\rangle = 2D\delta^{d-1}(\vec{x}-
\vec{x}')\delta  (t-t')$, where $D$ is a constant, and the brackets
denote a
statistical
average.  The vector $\vec x$ determines positions in a
$(d-1)$-dimensional plane \cite{Dimension} of a full space ${\vec r} =
({\vec x} , y)$, while $t$ is time.  The term proportional to $\lambda$
describes lateral growth of the interface, and plays an important role
in roughening. This nonlinearity also implies the
model has no free energy.

Salient features of this system are the average
velocity $\partial \langle h \rangle /\partial t \sim {\rm const.}$,
and the interface width
$W \equiv \sqrt{\overline{h^2}- (\overline{h})^2 }$, where the bar
denotes spatial average.
The width obeys
\begin{equation}
\label{scaling}
W(L,t)\sim L^{\chi} f(tL^{-z}),
\end{equation}
for late times and large length scales, where $L$ is the linear size of
the growing substrate, and $f$ is a scaling function.
In any dimension,
Eq. (1) presents an invariance, commonly referred to as ``Galilean'',
providing an exact relationship between $\chi$ and $z$. In addition, for
$d=2$, a fluctuation-dissipation theorem allows one to calculate the interface
exponents, $\chi=1/2$ and $z=3/2$ \cite{Forster}.
These are
consistent with numerical simulations and differ from the trivial
exponents obtained for $\lambda=0$:  $\chi=(3-d)/2$ and $z=2$
\cite{Edwards}.
A renormalization-group analysis shows that the critical dimension of
the model is $d_c = 3$, but that it involves an infra-red unstable
fixed point.  Thus $d_c$ is thought to be a lower critical dimension
for a transition which occurs in $d>3$
between a strong coupling and a weak coupling behavior.

Many justifications of this equation have been made in different
contexts, particularly for sputtered growth of epitaxial layers.  Here
we focus on the interface of a stable phase growing
at the expense of
a metastable
phase \cite{GGG}.  One example of such a system is the full field model
of critical dynamics, called model A \cite{Hohenberg}.  There a nonconserved
order parameter field $\psi({\vec r}, t)$ is the only dynamical mode.  In
equilibrium, at low enough temperatures, a stable interface exists
between coexisting
uniform
phases $\psi \approx \pm 1$.  If a system is
prepared in this fashion and a constant external field $H$ which is
conjugate to $\psi$ is applied, one of the two phases becomes
metastable, and the interface
advances through it.
It can be shown that the dynamics
of the interface separating those phases is described by the KPZ
equation, with $\lambda \propto H$ \cite{Weaseltalk}.  Systems such as
the surfaces of growing dendrites also correspond to the KPZ equation
if one takes the appropriate limit of model C, where in addition to
$\psi$, there is a conserved field coupled asymmetrically to the order
parameter. In this case, the diffusion length of the conserved field must
be much larger than the system size.

The purpose of this Letter is to study the interesting possibility of a
{\it time varying} external field $H(t)$ in model A. The mapping giving the
KPZ equation still follows, but now with
\begin{equation}
      \lambda(t) \propto H(t).
\end{equation}
Since $H$ is an external field its time dependence can be prescribed in
any fashion. Most importantly, we expect it to be experimentally tunable, as we
shall discuss below.

Assuming now a time-dependent $\lambda(t)$ in the KPZ equation (\ref{kpz}),
there are a number of interesting properties, three of
which we now describe. Our main working hypothesis will be that the
scaling relation (\ref{scaling}) still remains valid.
\begin{itemize}

\item[1.]
The equation has an invariance under
a generalized
``Galilean'' transformation
\begin{equation}
{\vec x} \rightarrow
{\vec x} - {\vec \epsilon} \int^t_{{\rm const.}} \lambda (t')\, dt',
\; \; {\rm and} \; \; h \rightarrow h +  {\vec \epsilon} \cdot {\vec x},
\end{equation}
where ${\vec \epsilon}$ is a constant vector of infinitesimal
amplitude.  This implies that $\lambda(t)$ is not renormalized
under the application of the renormalization group,
since it is directly involved in an exact invariance of the equation.  Here
we consider a power-law dependence of $\lambda(t)$ on time as:
\begin{equation}
     \lambda(t) = \frac{\lambda_0}{t^{\alpha}},
\end{equation}
where $ \lambda_0$ is a constant, and report results for $\alpha > 0$
(we have also studied $\alpha < 0$). The invariance above implies the novel
relations:
\begin{equation}
\label{chipz}
 \chi + z = 2 +z\alpha, \; \; {\rm and} \; \; d_c = 3-4\alpha,
\end{equation}
where the {\it tunable critical dimension} is obtained from the trivial
$\lambda = 0$ exponents.
If the scaling hypothesis is valid, these are exact results.
We note that in the asymptotic
regime, $\partial \langle h (t) \rangle/\partial t \sim t^{-\alpha}$.
This model has been studied previously by Lipowsky \cite{Lipowsky} in a
different context.

\item[2.]
The importance of the result (6) lies in that a nonzero
$\alpha$ could be sufficient to
move to lower dimensions the phase transition which the usual KPZ equation
exhibits only for $d>3$. This will make it experimentally accessible.
For such a
transition, at small $\lambda_0$ we should have the trivial exponents,
which satisfy $(2-\chi )/z = {\rm min}((d+1)/4,1)$ , while beyond
some larger value of $\lambda_0$, (\ref{chipz}) should hold, implying
$(2-\chi )/z = 1-\alpha$. In
principle, it should also be possible to investigate this transition
analytically, but we have not been able to do so.
We intend to numerically study the possibility of this transition in the
future.

\item[3.]
The nontrivial requirement of Galilean invariance implies that
exponents are tunable in the presence of a time-varying field.  In
$d=2$, a fluctuation-dissipation relation is satisfied both for $\lambda =0$
and $\lambda = {\rm const.\ }$ (i.e., for $\alpha = +\infty$ and
$\alpha = 0$) giving $\chi =1/2$.  We therefore conjecture that the value
of $\chi$ remains $1/2$ for any intermediate
$\lambda \propto 1/t^{\alpha}$, with $\alpha > 0$. This hypothesis and
Eq.(\ref{chipz}) fix the values of the scaling exponents:
\begin{equation}
\chi = \frac{1}{2}, \; \; {\rm and}  \; \; z = \frac{3}{2(1-\alpha)}
\end{equation}
for $d=2$, with $0\le \alpha \le 1/4$.  For $\alpha >
1/4$ one is above the critical dimension so that, at least for
small $\lambda_0$,
we expect the ideal interface results with
$z=2$.  It then follows that one could have a phase
transition in two dimensions for $\alpha > 1/4$, if $d_c$
remains a lower critical dimension.  Of course, the fixed point could
become infra-red stable for some $\alpha$ in $0< \alpha < 1/4$, so that
there would be no transition in $d=2$.
In $d=3$, the situation may be even more complicated.
If $\chi$ were to take the ideal interface value of 0 for $\alpha > 0$,
with $z$
determined from Eq. (\ref{chipz}), the scaling exponents would jump
discontinuously for $\alpha > 0$. This question will be studied in
more detail in future work.

\end{itemize}

We have
checked our hypothesis
numerically with two models for $d=2$.  First, we have directly
integrated the KPZ equation with a time-dependent $\lambda$.
Second, we have simulated the restricted solid-on-solid growth (GRSOS) model
\cite{Kim} using a time-varying growth rate.

The details of our numerical work are as follows.
The KPZ equation, discretized as in \cite{GGG} was
integrated by the Euler method. Systems sizes ranging from $50$ to
$50 000$ lattice sites were studied. The time step was generally taken as
$0.02$, and some smaller values were considered to check accuracy.
Typical values for the rest of the parameters were $\nu=3.5$, $\lambda_0=50$
and $D=0.01$, and several values of $\alpha$ were considered.
For the GRSOS model, systems of sizes up to $L=50 000$ were simulated also.
The algebraic decay of $\lambda$ was achieved by incorporating time-dependent
condensation {\it and} evaporation rates, whose difference is proportional
to $\lambda$. For both models, the known limits for $\alpha=0$ and
$\alpha=\infty$ were checked.

Direct determinations of $\chi$ were obtained by running systems of several
sizes $L$ until
saturation, and then using the scaling of $W(L,t=\infty)$ with $L$.
For both models, we obtained $\chi \approx
0.5$ for all values of $\alpha$ studied, as shown in Fig. 1.
Our conjectures are then most simply tested through
the exponent $\beta =\chi /z$, where $W \sim t^{\beta}$ if $t<<L^z$.
For $d=2$ we expect
\begin{equation}
\label{beta}
 \beta = \frac{1-\alpha}{3}
\end{equation}
for $0\le \alpha \le 1/4$ and $\beta = 1/4$ for $\alpha
\ge 1/4$. To obtain $\beta$, we performed extensive calculations
of the time-dependence of the width. Our results
are summarized in Fig. 2. Although it becomes increasingly difficult to
extract $\beta$ around $\alpha= 1/4$, our results for various values
of $\alpha$ are in complete agreement with Eq. (\ref{beta}).

A somewhat different way of checking our prediction for the exponents
can be obtained from the {\it crossover scaling form} of $W(L=\infty,t)$
proposed in \cite{GGG}, which for $d=2$ reads:
\begin{equation}
\label{crossover}
W(L=\infty,t) \sim t^{\frac{1}{4}} g(t \lambda^\phi) \ ,
\end{equation}
with $g(x) \sim x^{1/12}$ for small $x$. Theoretical arguments and
computer simulations of discrete growth models \cite{phi4}
have so far given $\phi=4$, while direct simulations of Eq. (\ref{kpz})
seem to give \cite{GGG,phi3} $\phi=3$. If we assume that
(\ref{crossover}) remains valid for $\lambda(t) \sim t^{-\alpha}$, we find
$W \sim t^{(4-\alpha \phi)/12}$, if $\alpha < 1/\phi$. For $\phi=4$ this
agrees with our $\beta$ in $d=2$. In fact, we have checked the scaling
form (\ref{crossover}), with a time-dependent $\lambda$ for {\sl both}
 the GRSOS
model {\sl and} the KPZ
equation, and obtained $\phi=4$ for both models. This lends strong
support to our direct determination of $\beta$ (see Fig. 2).
To obtain this result for the KPZ equation, it is crucial to
realize that the lattice constant
of the mesh in which we solve the equation must be smaller than the length
scale $\nu^3/\lambda^2 D$. For coarser discrete meshes, as used in the
previous works \cite{GGG,phi3}, we observe a
crossover to $\phi=3$. Details of these results will be published elsewhere.

Experimentally, the phenomena described here can be probed with
metastable systems such as a solid growing into a supercooled melt or
into a supersaturated solution.  The degree of metastability, which is
proportional to $\lambda$, can be varied by, for example, progressively
decreasing the degree of undercooling.  Similarly, in a sputtered
growth experiment, our results describe a system as the sputtering rate
is tuned to fall off algebraically in time.  Thus we expect it to be
experimentally feasible to probe the three properties we have
discussed: The generalized relation of Galilean invariance, the possibility
of a phase transition at physical dimensions, and the tunable
roughening exponents.

\bigskip

This work was supported by the Natural Sciences and Engineering
Research Council of Canada, and {\it le Fonds pour la Formation de
Chercheurs et l'Aide \`a la Recherche de la Province de Qu\'ebec\/}.
E.H-G acknowledges support from {\it Comisi\'on Interministerial de Ciencia y
Tecnolog\'\i a} (Spain). T.A-N. was also supported by the Academy of Finland,
and acknowledges the hospitality
of McGill University during his visit there.

\newpage

{\Large \bf Figure Captions}

\begin{itemize}

\item[1.]
The exponent $\chi$ as a function of $\alpha$ from simulations of the KPZ
equation (diamonds) and the GRSOS model (asterisks). KPZ data come from
averages over 100 realizations of systems of sizes $L=50$, $100$, $200$, and
$400$. GRSOS data have been averaged over $10^7-10^8$ Monte Carlo steps for
$L=100$, $200$, $500$ and $1000$. The error bars are purely statistical;
no systematic finite size errors have been included.

\item[2.]
The exponent $\beta$ as a function of $\alpha$. Meaning of the symbols as in
Fig. 1. Solid line is the prediction of Eq. (\ref{beta}). Data come from
simulations of systems of size $L=50 \ 000$ averaged over 30 (KPZ) and 30-60
(GRSOS) independent runs.

\end{itemize}


\begin{thebibliography}{99}
\bibitem{Dimension}
Here, $d$ is the dimension of the full space.  A common notation is to
write $d = (d-1) +1$.  In our notation $1+1 = 2$, $2+1 = 3$, and so
on.
\bibitem{Weaseltalk}
The mapping follows on large time scales which are, nevertheless,
sufficiently small that nucleation does not take place in the
metastable state.
\end{thebibliography}
\end{document}